\documentclass[amsmath,amssymb, aps, pra, twocolumn,superscriptaddress]{revtex4-2}

\usepackage{float}
\usepackage{graphicx}
\usepackage{dcolumn}
\usepackage{bm}
\usepackage{hyperref}
\hypersetup{
	colorlinks=true,
	linkcolor=blue,
	filecolor=magenta,      
	urlcolor=cyan,
}
\usepackage{color}
\usepackage{fancyhdr}
\usepackage[normalem]{ulem}
\usepackage{physics}
\usepackage[dvipsnames]{xcolor}
\usepackage{upgreek}  

\definecolor{greenishred}{rgb}{0.6, 0.4, 0}

\begin{document}
	
	\title{Thermalization of a Trapped Single Atom with an Atomic Thermal~Bath}
	
	\author{Rahul Sawant}
	\email{r.v.sawant@bham.ac.uk}
	\affiliation{School of Physics and Astronomy, University of Birmingham, Edgbaston, Birmingham, B15 2TT, UK}
	\author{Anna Maffei}
	\affiliation{School of Physics and Astronomy, University of Birmingham, Edgbaston, Birmingham, B15 2TT, UK}
	\affiliation{Dipartimento di Fisica “E. Fermi”, Universit\`a di Pisa, Largo B. Pontecorvo 3, 56127 Pisa, Italy}
	\author{Giovanni Barontini}
	\affiliation{School of Physics and Astronomy, University of Birmingham, Edgbaston, Birmingham, B15 2TT, UK}
	
	\begin{abstract}
		We studied a single atom trapped in an optical tweezer interacting with a thermal bath of ultracold atoms of a different species. Because of the collisions between the trapped atom and the bath atoms, the trapped atom undergoes changes in its vibrational states occupation to reach thermal equilibrium with the bath. By using Monte Carlo simulations, we characterized the single atom's thermalization process, and we studied how this can be used for cooling. Our simulations demonstrate that, within known experimental limitations, it is feasible to cool a trapped single atom with a thermal bath. 
	\end{abstract}
	\maketitle

\section{Introduction}
One of the current goals in physics is to fully understand the thermodynamics of quantum systems~\cite{Sai_2016_QT}. The understanding of how open quantum systems behave thermodynamically is especially important, as it could be the key to unveil how classical mechanics emerges from the quantum, and will help the development of quantum technologies. The paradigmatic model of open quantum systems is the \emph{spin-boson model}~\cite{Leggett_Dynamics_1987,Weiss_book_2012}, i.e., a system in which a single two-level particle interacts with a thermal bath of harmonic oscillators. In this work, we considered an experimentally implementable extension of such a celebrated model: a single atom trapped in a harmonic potential interacting with a bath of ultracold atoms of a different species.

Ultracold atoms are highly controllable and versatile systems that can be used to simulate a large variety of quantum phenomena~\cite{Georgescu_2014, Schafer2020}. 
Optical tweezers are emerging as one of the most promising and powerful tools for quantum simulations with cold \mbox{atoms~\cite{Kaufman2014,Endres_2016,deLeseleuc2019}}, and have been recently proposed as a platform to study quantum thermodynamics and implement single atom engines~\cite{Barontini_2019}. 
In the latter system, a single atom interacts with a thermal bath and undergoes various thermodynamic cycles. One or more of the strokes of these engines involves the thermalization of the single atom with a bath of atoms of a different species. With an eye to such quantum engines, we simulated the interaction of a single atom in an optical tweezer with a thermal gas of atoms trapped in a macroscopic trap. The thermalization process studied here could also be used as an alternative method to cool single trapped atoms and bring them to the ground state of their tweezer potential, which can currently be accomplished only with laser cooling~\cite{Kaufman_2012}.

The article is organized as follows. In Section~\ref{sec:theory}, we describe our system and provide the theory that we have used to simulate our system. In Section~\ref{sec:sym_cooling}, we provide the results of a Monte Carlo simulation where we show how the thermalization rates vary and how the trapped atom interacts with a bath of atoms undergoing evaporative cooling. Section~\ref{sec4} is devoted to the conclusions.
\section{The System}
\label{sec:theory}

\begin{figure}[b]
    \centering
    \includegraphics[width=0.35\textwidth]{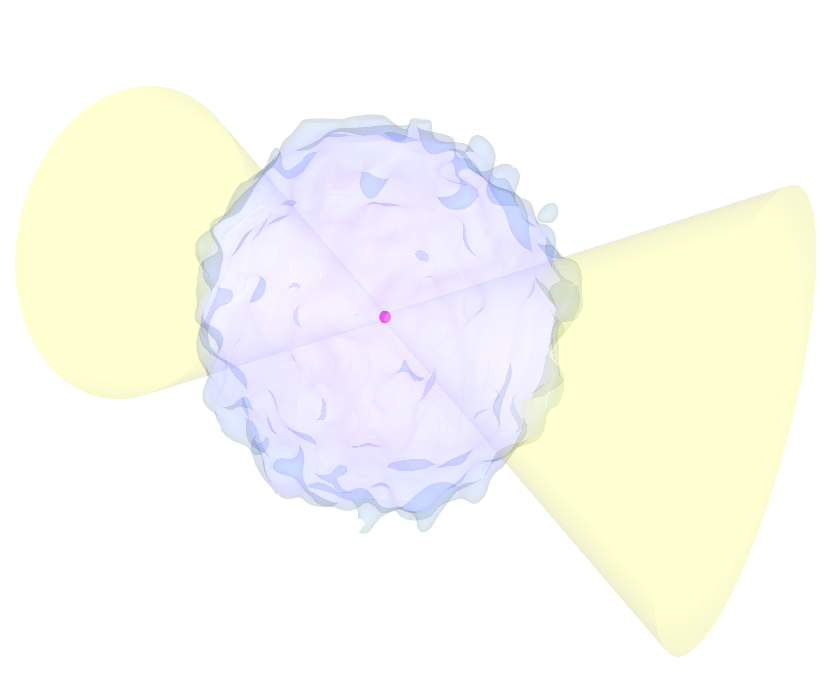}
    \caption{Pictorial representation (not to scale) of the system under study: a single atom (red sphere) is trapped in an optical tweezer (yellow beam) and is immersed in a bath of atoms of a different species (blue cloud).}
    \label{fig:cartoon}
\end{figure}

The system under study is pictorially represented in Figure~\ref{fig:cartoon}. A single atom is trapped in an optical tweezer and interacts with a cloud of thermal atoms. Our aim was characterizing how the interaction with the bath leads to the thermalization of the single atom, meaning that it reaches the Gibbs state $G(T)=1/Z \sum_ne^{-E_n/k_B T}$, with $Z$ the partition function, $E_n$ the energy of the n$^\text{th}$ level, $k_B$ the Boltzmann's constant and $T$ the temperature of the bath. Therefore, in the following, we refer to the the \emph{temperature of the single atom} $T_A$ as the temperature associated with $G(T_A)$. 

For a single atom in an optical tweezer, when the trap depth is significantly greater than the temperature of the atom, we can approximate the tweezer as a three-dimensional harmonic trap. Since optical tweezers are produced with a single beam, they are characterized by a strong anisotropy, with the axial direction of propagation of the beam providing a weaker confinement than the radial directions. In addition, we consider the bath as made of thermal bosonic atoms. After a collision with the bath atom, the trapped atom can undergo a change in the vibrational level it occupies. Which level it goes to depends on the overlap of the initial and the final wavefunctions. If the atom changes its vibrational state from $\boldsymbol{n} = (n_x,n_y,n_z)$ to $\boldsymbol{m} = (m_x,m_y,m_z)$, the relative rate is given by~\cite{Zoller1995,Papenbrock2002},
\begin{widetext}
\begin{equation}
    \gamma_{\boldsymbol{n},\boldsymbol{m}} = C \int dt \int d\vec{k} d\vec{k}' n(\vec{k}) (n(\vec{k'})+1) |T(\boldsymbol{n},\boldsymbol{m},k,k')|^2 \\
    e^{\left(\frac{k^2}{2 m_B}-\frac{k'^2}{2 m_B} + \alpha \hbar \omega\right)t}.
\end{equation}
\end{widetext}
where $C$ is a constant common to all the rates, $\vec{k}$ and $\vec{k}'$ are the wavevectors of the bath atom before and the after the collision, $m_B$ is the mass of the bath atom, $\omega$ is the tweezer radial trapping frequency, $\alpha = (n_x - m_x) + \eta_y (n_y - m_y) + \eta_z (n_z - m_z)$, $\eta_y/\eta_z$ is the aspect ratio of the trap, $n(\vec{k})$ is the number of bath atoms with wave-vector $\vec{k}$ , and $T(\boldsymbol{n},\boldsymbol{m},k,k')$ is the overlap between the initial and the final wavefunctions of both the trapped atom and the bath atom. In Equation (1), it is assumed that the bath atom is a free particle during the collision. The above integrals for an isotropic trap were solved in References~\cite{Zoller1995,Papenbrock2002}, including the trap anisotropy, and we find: 
\begin{widetext}
\begin{equation}
    \gamma_{\boldsymbol{n},\boldsymbol{m}}  =  \tilde{C} e^{1/2 \alpha' \delta} \sum_{l_x = 0, k_x = 0}^{p_x} \sum_{l_y = 0, k_y = 0}^{p_y} \sum_{l_z = 0, k_z = 0}^{p_z} \left[\prod_{j = x,y,z} C_{m_j,n_j,l_j} C_{m_j,n_j,k_j} \Gamma\left({q_j  + 1/2}\right)  V(\alpha',\delta)_{1+q_x+q_y+q_z} \right],
\end{equation}
\end{widetext}
where $\tilde{C}$ is a constant common to all rates, $\alpha' = \alpha m_B/m_S$, with $m_S$ the mass of the trapped atom, 
\begin{equation}
\delta = \frac{m_S}{m_B} \frac{\hbar \omega (1+\eta_y + \eta_z)}{k_B T_B}, 
\end{equation}
where $T_B$ is the bath temperature, $\Gamma$ is the Gamma function and $q_j = m_j  + n_j - l_j - k_j$, $p_j$ = min$(m_j,n_j)$, 
\begin{equation}
    C_{m,n,l} = \frac{(-1)^l \sqrt{m!n!}}{l!(m-l)!(n-l)!},
\end{equation}
and
\begin{equation}
\begin{split}
    V(\alpha',\delta)_{i} &= 2 \sqrt{\frac{\delta}{\pi}} \Gamma(1/2- i) \left(\frac{-|\alpha'|\delta}{2\sqrt{\delta (1+\delta/4)}}\right)^i \\
    & \times K_i\left(|\alpha'|\sqrt{\delta (1+\delta/4)}\right).
\end{split}
\end{equation}
{$K_i(x)$ denotes the modified Bessel function and $i$ is a positive integer.} Once we calculate the above rates, at each collision, the probability to jump from state $\boldsymbol{n}$ to $\boldsymbol{m}$ can be calculated~as 
\begin{equation}
p_{\boldsymbol{n}\rightarrow\boldsymbol{m}} = \frac{\gamma_{\boldsymbol{n}, \boldsymbol{m}}}{\sum_{\boldsymbol{v}}\gamma_{\boldsymbol{n},\boldsymbol{v}}}.
\end{equation}
In the above calculations, we assume that the trapped atom interacts with a uniform density bath. This is a valid assumption for systems where the trapped atom is confined to a smaller volume compared to the trap of the bath atoms. 

To account for the finite trap depth, we restrict our calculations to a suitably chosen $(n_x^\text{max},n_y^\text{max},n_z^\text{max})$ such that $ n_x^\text{max}, \eta_y n_y^\text{max}, \eta_z n_z^\text{max}  \gg k_B T_A/(\hbar \omega)$. Here, $T_A$ is the initial temperature of the single atom. For the simulations in this article, we first calculate the $p_{\boldsymbol{n}\rightarrow\boldsymbol{m}}$ for all the level combinations for each bath temperature $T_B$. The second step is to perform Monte Carlo simulations where the atom jumps between the levels according to the probabilities $p_{\boldsymbol{n}\rightarrow\boldsymbol{m}}$. 

\section{Results}
\label{sec:sym_cooling}
\begin{figure}[h]
    \centering
    \includegraphics[width=0.49\textwidth]{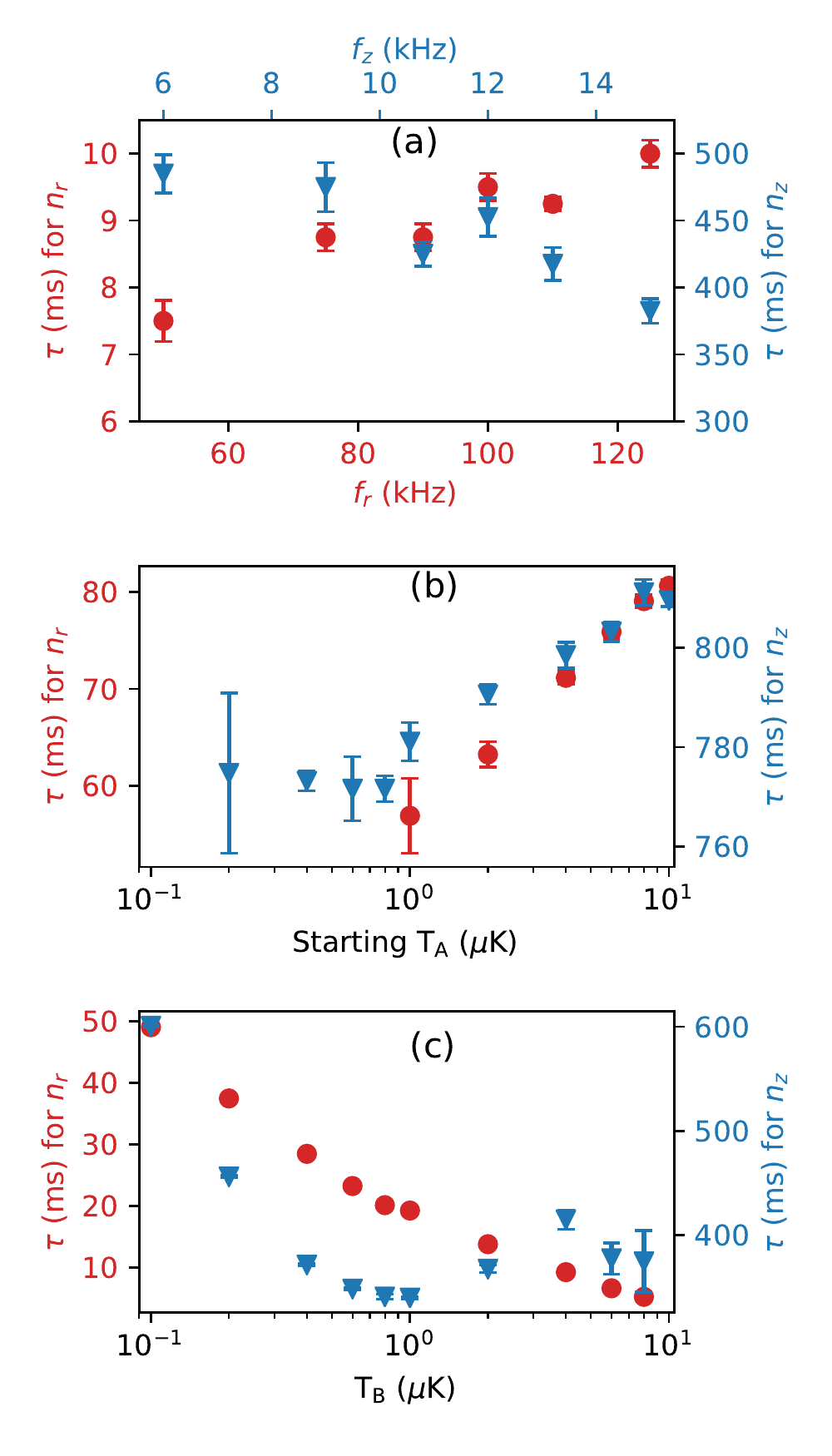}
    \caption{The variation of thermalization time constants of radial and axial vibrational levels of the trapped atom with respect to (\textbf{a}) the trapping frequency, (\textbf{b}) the starting temperature of the trapped atom, and (\textbf{c}) temperature of the bath. The red circles are for radial direction and the blue triangles are for the axial direction. Here $a = 160 a_0$, $n_B = 7.7 \times 10^{12}$~cm$^{-3}$, $\omega/(2 \pi) = 100$~kHz for (\textbf{b},\textbf{c}), $T_B = 4~\upmu$k, and $T_B = 0.1~\upmu$k for (\textbf{a},\textbf{b}) respectively. {$T_A = 8~\upmu$k and $T_A = 10~\upmu$k for (\textbf{a},\textbf{c}) respectively.} For all the data points $\eta_y = 0.98$ and $\eta_z = 0.105$. }
    \label{fig:tau_vs_all}
\end{figure}

We first explore how the rate of thermalization for the trapped atom changes with respect to various parameters. Second, we simulate what happens when the trapped atom is interacting with a bath  undergoing evaporative cooling. To illustrate a specific case, in our simulations we use $^{87}$ Rb and $^{41}$ K as the bath atoms and the single atom, respectively, similar to what was done in~\cite{Barontini_2019}. 

We are interested in the rate at which the trapped atom thermalizes with the bath atoms. We ran the Monte Carlo simulations to know how many \emph{elastic} collisions it takes for the radial and axial vibrational states of the trapped atom to thermalize with the bath. {To calculate the thermalization time constant, $\tau$, we multiplied the number of collisions with the time between collisions, which is given by 
\begin{equation}
    \label{eqn:elas_colls}
    \tau_\text{col} = 1/(4\pi  a^2 n_B \bar{v}),
\end{equation}
where $n_B$ is the density of bath atoms, $a$ is the heteronuclear s-wave 
scattering length between the bath and the trapped atom, 
$\bar{v} = \sqrt{8 k_B  T_B/(\pi \upmu)}$ is the mean velocity at bath temperature ($T_B$), and $\upmu$ is the reduced mass between the bath atom 
and the single atom. For the simulations, we consider bath temperatures in 
the range $0.1~\upmu$K-$10~\upmu$K, and the atomic density in the bath to 
be $\sim$$10^{12}$~cm$^{-3}$. 

These parameters can be produced in current experimental setups~\cite{Munoz_2020_dissipative}. For the interspecies scattering length that sets the strength of the interaction between the trapped atom and the bath, we have chosen $a = 160 a_0$, which can be achieved exploiting one of the KRb heteronuclear Feshbach resonances \cite{thalhammer}. Unless specified otherwise, we have set $\omega = 100$~kHz, $\eta_y = 0.98$ and $\eta_z = 0.105$, which are realistic values for tweezers experiments~\cite{Kaufman_2012}}. {As $\omega_x\simeq\omega_y$, from here, we refer to either of them as the radial frequency $\omega_r$ and we identify the radial direction with both $x$ and $y$.} The results of our simulations for various parameters are shown in Figure~\ref{fig:tau_vs_all}, where we report mean values and standard deviations of the mean of $\tau$ obtained with 6 repetitions of 5000~trajectories each. 
\begin{figure}[b]
    \centering
    \includegraphics[width=0.49\textwidth]{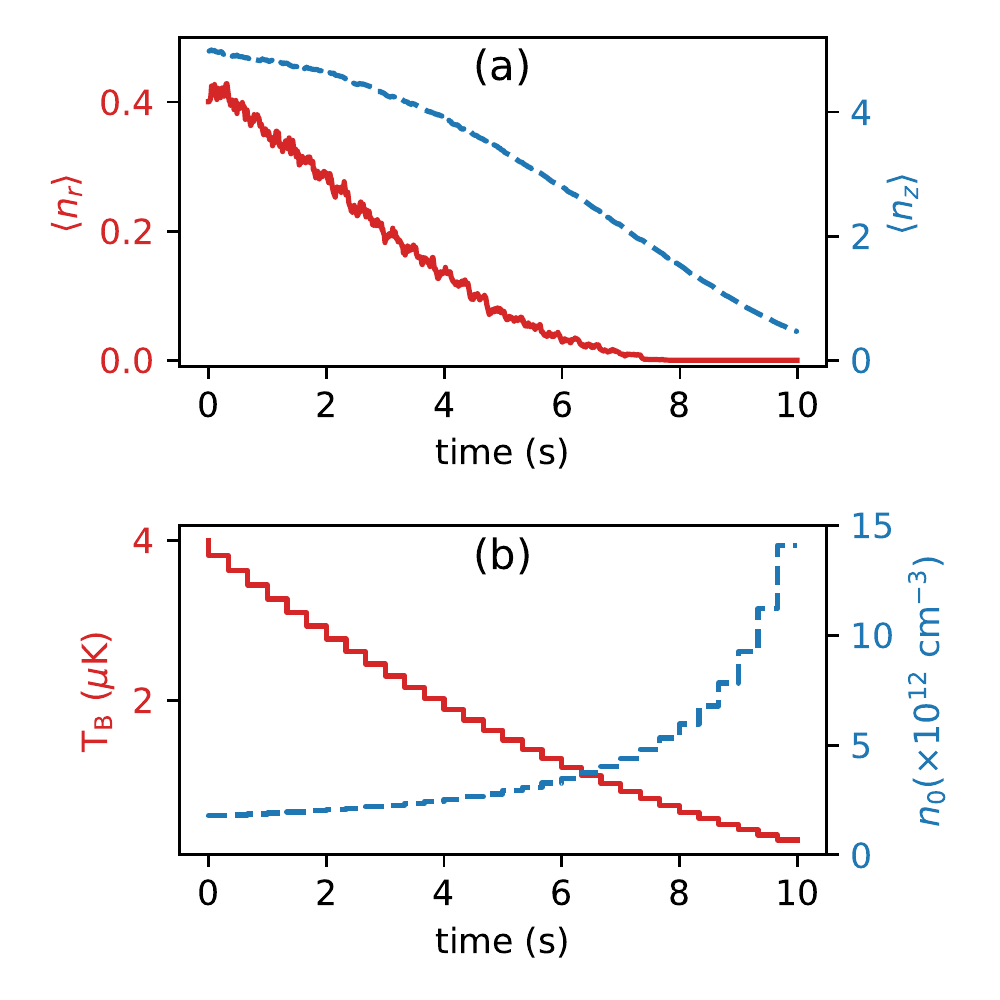}
    \caption{Trapped atom interacting with a bath that is continuously undergoing evaporative cooling. (\textbf{a}) Radial and axial mean vibrational occupation numbers are shown by red solid and blue dashed lines, respectively. {Here, the final occupation numbers are $\langle n^{\prime}_r \rangle = 0.00$ and  $\langle n^{\prime}_z \rangle = 0.46$, meaning that the atom always ends up in the ground state of the radial direction. (\textbf{b}) The temperature and the atomic density of the bath atoms are shown by red solid and blue dashed lines, respectively. Here, $\omega/(2 \pi) = 100$~kHz, $a = 160 a_0$, $\eta_y = 0.98$, and $\eta_z = 0.105$ for all graphs.}}
    \label{fig:evap_long}
\end{figure}

\begin{figure}[b]
    \centering
    \includegraphics[width=0.49\textwidth]{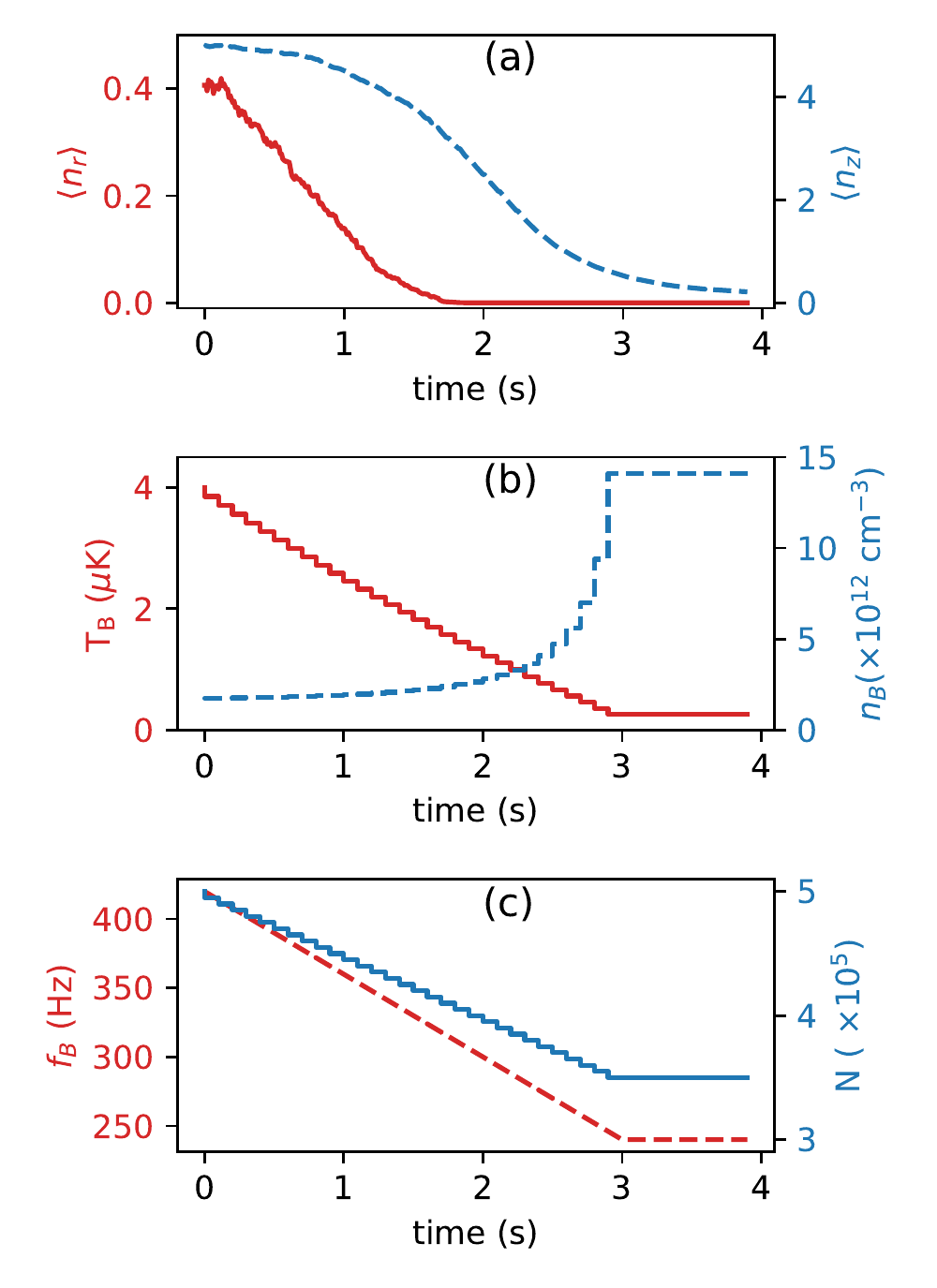}
    \caption{Trapped atom interacting with a bath that is undergoing fast evaporative cooling for three seconds. (\textbf{a}) Radial and axial mean vibrational occupation numbers are shown by red solid and blue dashed lines, respectively. {Here, the final occupation numbers are $\langle n^{\prime}_r \rangle = 0.00$ and  $\langle n^{\prime}_z \rangle = 0.23$, meaning that the atom  always ends up in the ground state of the radial direction.} (\textbf{b}) The temperature and the atomic density of the bath atoms are shown by red solid and blue dashed lines, respectively. {(\textbf{c}) The red dashed line shows the change in the trapping frequency of the bath ($f_B$) and the blue solid line shows the variation in the bath atom number.} Here, $\omega/(2 \pi) = 100$~kHz, $a = 160 a_0$, $\eta_y = 0.98$, and $\eta_z = 0.105$ for all graphs. }
    \label{fig:evap}
\end{figure}

As expected, from Figure~\ref{fig:tau_vs_all}a we can note that, because the trapping frequency is low in the axial direction (denoted by $z$), the time required for thermalization is higher than in the radial direction (denoted by $r$). 
The thermalization time does not show a strong dependence on the trapping frequency within the parameters of interest because the conditions $\hbar \omega  \sim k_B T_B$ and $ \hbar \eta_z \omega \ll k_B T_B $ do not change substantially for the radial and the axial directions, respectively. In Figure~\ref{fig:tau_vs_all}b, we keep $T_B$ constant at $0.1~\upmu$k and vary the starting temperature of the trapped atom, $T_A$. Here,  $\tau$ increases for both directions as the initial $T_A$ increases, but the increase is rather contained. {For this figure panel, there are no datapoints for radial direction below $T_A= 1~\upmu$k as the probability to be in $n_x=n_y = 0$ state is practically $100\%$ and, hence, further cooling is not possible. 
Figure~\ref{fig:tau_vs_all}c shows the variation of the thermalization time with respect to the bath temperature, where $T_A = 10~\upmu$k for all points.
Here we find that, as the bath temperature is reduced, the thermalization time required for the radial direction increases. This is because, according to\mbox{ Equation~(\ref{eqn:elas_colls})}, the time between collisions increases when the bath temperature decreases. The axial direction does not show such a simple behavior, the thermalization time remains initially fairly constant until $T_B\simeq1~\upmu$k, when it has a minimum, and then increases below such temperature. This is because although the time between collisions increases, the number of collisions required to cool the single atom reduces as $p_{\boldsymbol{m}\rightarrow\boldsymbol{n}}/p_{\boldsymbol{n}\rightarrow\boldsymbol{m}}$ increases, where $\boldsymbol{n} < \boldsymbol{m}$.   
The above results tell us that (as expected) the thermalization of the axial direction is by far the limiting factor in all the conditions we have explored. Additionally, we have found that in order to effectively cool the trapped atom, there exists an optimal bath temperature that minimizes the axial thermalization time and therefore could be used to speed up the entire thermalization process.}
\begin{figure}[b]
    \centering
    \includegraphics[width=0.49\textwidth]{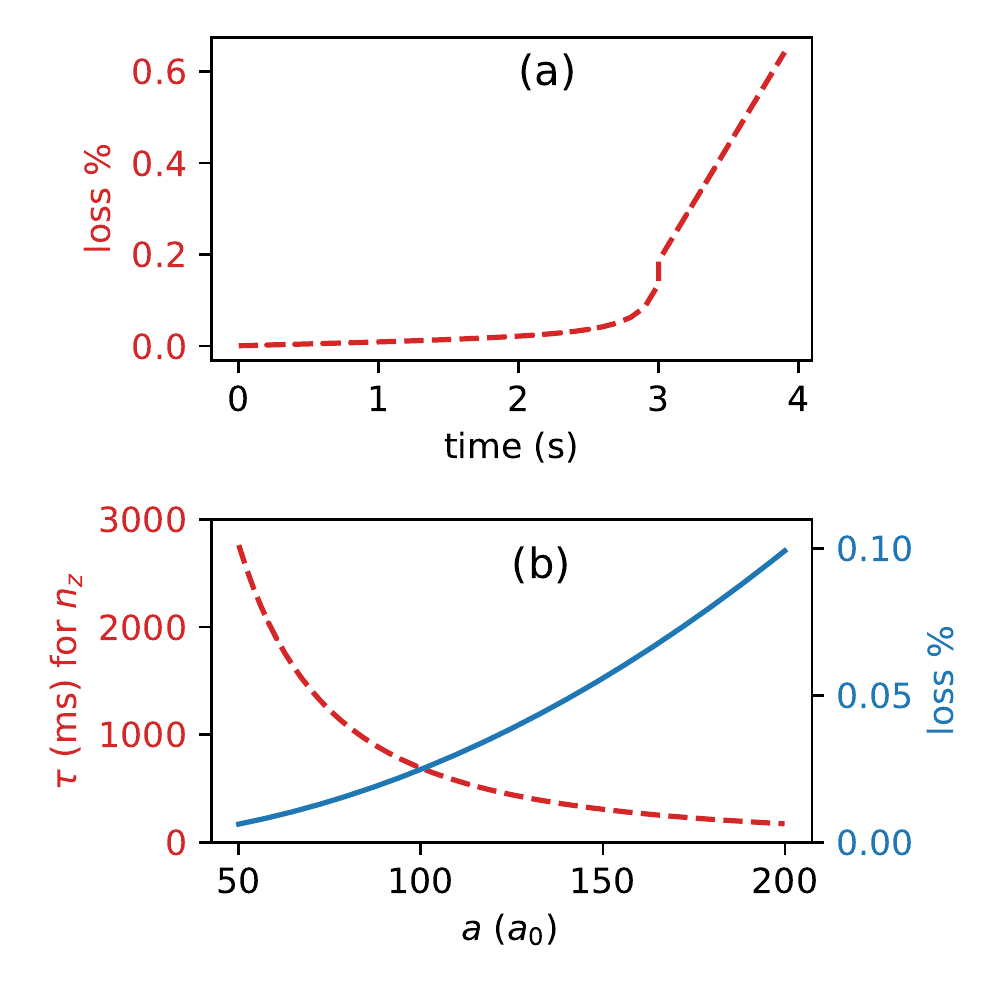}
    \caption{{Three-body loss during thermalization. (\textbf{a}) The red dashed line shows the percentage of times the trapped atom will be lost due to three-body collisions during the evaporative cooling shown in Figure~\ref{fig:evap}. (\textbf{b}) Effect of change in scattering length on (i) thermalization time (red dashed) and (ii) the percentage of times the single atom will be lost due to a three-body collision during one thermalization period (blue solid). Here, $T_B = 1~\upmu$k and $n_B = 10\times10^{12}$~cm$^{-3}$ for (b).}}
    \label{fig:three-body}
\end{figure}

In Figure~\ref{fig:evap_long} we show the results of a Monte Carlo simulation where the trapped atom is immersed in a bath whose temperature is decreasing due to evaporative cooling. In evaporative cooling, the temperature, the number of atoms, and the trapping frequency of the bath atoms reduce continuously. As we need to calculate $p_{\boldsymbol{n}\rightarrow\boldsymbol{m}}$ for each $T_B$, we consider the change in these parameters in step-wise manner. The figure shows how the average vibrational state of the trapped atom changes as we move along the evaporation ramp. The length of the ramp has been chosen to ensure $T_A\simeq T_B$ at every time. An alternative, faster approach is shown in Figure~\ref{fig:evap}a,b. In this case, the radial vibration levels are able to follow the bath temperature because of the fast thermalization with the bath. The axial direction is always lagging behind as the time taken for the thermalization of the axial direction is higher compared to timescales of evaporation. In order for the axial direction to catch up, we then add a waiting time. This allows enough time for all the degrees of freedom of the trapped atom to thermalize with the bath atoms and reach $T_A=T_B$. Notably, with this second method, we are able to sensibly reduce the evaporation time compared to the first~ramp. 

To further speed up the thermalization, i.e., to increase the axial direction's thermalization rate, one could either increase the density of the bath or increase the scattering length. Both of these strategies also lead  to an increase in the harmful three-body collision rate, as shown in Figure~\ref{fig:three-body}a,b. We convert the three-body loss rate to total probability of loss during the evaporation of the bath for our case, which is shown in Figure~\ref{fig:three-body}a. The three-body loss rate here is calculated as $\tau_{loss}=4 \pi \hbar a^4 n_B^2/m_B$~\cite{Fedichev_1996}. In Figure~\ref{fig:three-body}b, we show the effect of change in scattering length to $\tau$ and the loss probability, $1-e^{-\tau/\tau_{loss}}$, due to three-body collisions during thermalization time.

\section{Conclusions}
\label{sec4}
In conclusion, we investigated the thermalization of an atom trapped in an optical tweezer interacting with a thermal gas of ultracold atoms of a different species. We discussed the theory we used to understand how collisions between a trapped atom and the bath atoms translate into changes in the trapped atom's vibrational levels. With the help of a Monte Carlo simulation, we have shown how these changes lead to the trapped atom's cooling if the bath is at a lower temperature compared to the trapped atom. For parameters of practical interest, we found that the cooling rate in the radial direction is two orders of magnitude faster than the cooling rate in the axial direction. We have also shown the dependence of these cooling rates on the bath temperature, the tweezer's trapping frequency, and the single atom's starting temperature. Finally, we simulated a bath undergoing evaporative cooling and demonstrated that the trapped atom can be sympathetically cooled. We have shown that this could be an efficient scheme for cooling the single trapped atom, also considering three-body losses that limit the final efficiency. 

Besides the implementation of single atom quantum engines, this kind of system is a promising approach to studying out-of-equilibrium physics~\cite{Munoz_2020_dissipative}. The system studied here could be modified to obtain a pristine realization of several instances of the spin-boson model or other open quantum system models like the Caldeira--Leggett model~\cite{Caldeira_1981}. There are several additional exciting avenues that could be explored as extensions of this work. One could be to investigate the thermalization process in non-Markovian regimes, where the number of bath atoms is small or the scattering length is large. {Another could be to increase the thermalization times inspired by the recents works in References~\cite{Dann_shortcut_2019, Pancotti_speedup_2020}.} Alternatively, one could study a system in which the bath is realized with a pure Bose--Einstein condensate. The heat capacity of a Bose--Einstein condensate vanishes, making the feedback from the single atom significant. An exotic system could be realized with a partially condensed cloud, where two phases coexist and quantum and thermal fluctuations compete, {or by trapping a single ion in the tweezer.} Additionally, the use of squeezed states in the bath could lead to interesting quantum features that could boost the thermalization rate~\cite{Scully_2009,Dillenschneider_2009,Rossnagel_2014}.

\bibliography{bibliography}

\begin{thebibliography}{21}%
\makeatletter
\providecommand \@ifxundefined [1]{%
 \@ifx{#1\undefined}
}%
\providecommand \@ifnum [1]{%
 \ifnum #1\expandafter \@firstoftwo
 \else \expandafter \@secondoftwo
 \fi
}%
\providecommand \@ifx [1]{%
 \ifx #1\expandafter \@firstoftwo
 \else \expandafter \@secondoftwo
 \fi
}%
\providecommand \natexlab [1]{#1}%
\providecommand \enquote  [1]{``#1''}%
\providecommand \bibnamefont  [1]{#1}%
\providecommand \bibfnamefont [1]{#1}%
\providecommand \citenamefont [1]{#1}%
\providecommand \href@noop [0]{\@secondoftwo}%
\providecommand \href [0]{\begingroup \@sanitize@url \@href}%
\providecommand \@href[1]{\@@startlink{#1}\@@href}%
\providecommand \@@href[1]{\endgroup#1\@@endlink}%
\providecommand \@sanitize@url [0]{\catcode `\\12\catcode `\$12\catcode
  `\&12\catcode `\#12\catcode `\^12\catcode `\_12\catcode `\%12\relax}%
\providecommand \@@startlink[1]{}%
\providecommand \@@endlink[0]{}%
\providecommand \url  [0]{\begingroup\@sanitize@url \@url }%
\providecommand \@url [1]{\endgroup\@href {#1}{\urlprefix }}%
\providecommand \urlprefix  [0]{URL }%
\providecommand \Eprint [0]{\href }%
\providecommand \doibase [0]{https://doi.org/}%
\providecommand \selectlanguage [0]{\@gobble}%
\providecommand \bibinfo  [0]{\@secondoftwo}%
\providecommand \bibfield  [0]{\@secondoftwo}%
\providecommand \translation [1]{[#1]}%
\providecommand \BibitemOpen [0]{}%
\providecommand \bibitemStop [0]{}%
\providecommand \bibitemNoStop [0]{.\EOS\space}%
\providecommand \EOS [0]{\spacefactor3000\relax}%
\providecommand \BibitemShut  [1]{\csname bibitem#1\endcsname}%
\let\auto@bib@innerbib\@empty
\bibitem [{\citenamefont {Vinjanampathy}\ and\ \citenamefont
  {Anders}(2016)}]{Sai_2016_QT}%
  \BibitemOpen
  \bibfield  {author} {\bibinfo {author} {\bibfnamefont {S.}~\bibnamefont
  {Vinjanampathy}}\ and\ \bibinfo {author} {\bibfnamefont {J.}~\bibnamefont
  {Anders}},\ }\bibfield  {title} {\bibinfo {title} {Quantum thermodynamics},\
  }\href {https://doi.org/10.1080/00107514.2016.1201896} {\bibfield  {journal}
  {\bibinfo  {journal} {Contemporary Physics}\ }\textbf {\bibinfo {volume}
  {57}},\ \bibinfo {pages} {545} (\bibinfo {year} {2016})}\BibitemShut
  {NoStop}%
\bibitem [{\citenamefont {Leggett}\ \emph {et~al.}(1987)\citenamefont
  {Leggett}, \citenamefont {Chakravarty}, \citenamefont {Dorsey}, \citenamefont
  {Fisher}, \citenamefont {Garg},\ and\ \citenamefont
  {Zwerger}}]{Leggett_Dynamics_1987}%
  \BibitemOpen
  \bibfield  {author} {\bibinfo {author} {\bibfnamefont {A.~J.}\ \bibnamefont
  {Leggett}}, \bibinfo {author} {\bibfnamefont {S.}~\bibnamefont
  {Chakravarty}}, \bibinfo {author} {\bibfnamefont {A.~T.}\ \bibnamefont
  {Dorsey}}, \bibinfo {author} {\bibfnamefont {M.~P.~A.}\ \bibnamefont
  {Fisher}}, \bibinfo {author} {\bibfnamefont {A.}~\bibnamefont {Garg}},\ and\
  \bibinfo {author} {\bibfnamefont {W.}~\bibnamefont {Zwerger}},\ }\bibfield
  {title} {\bibinfo {title} {Dynamics of the dissipative two-state system},\
  }\href {https://doi.org/10.1103/RevModPhys.59.1} {\bibfield  {journal}
  {\bibinfo  {journal} {Rev. Mod. Phys.}\ }\textbf {\bibinfo {volume} {59}},\
  \bibinfo {pages} {1} (\bibinfo {year} {1987})}\BibitemShut {NoStop}%
\bibitem [{\citenamefont {Weiss}(2012)}]{Weiss_book_2012}%
  \BibitemOpen
  \bibfield  {author} {\bibinfo {author} {\bibfnamefont {U.}~\bibnamefont
  {Weiss}},\ }\href {https://doi.org/10.1142/8334} {\emph {\bibinfo {title}
  {Quantum Dissipative Systems}}},\ \bibinfo {edition} {4th}\ ed.\ (\bibinfo
  {publisher} {WORLD SCIENTIFIC},\ \bibinfo {year} {2012})\BibitemShut
  {NoStop}%
\bibitem [{\citenamefont {Georgescu}\ \emph {et~al.}(2014)\citenamefont
  {Georgescu}, \citenamefont {Ashhab},\ and\ \citenamefont
  {Nori}}]{Georgescu_2014}%
  \BibitemOpen
  \bibfield  {author} {\bibinfo {author} {\bibfnamefont {I.~M.}\ \bibnamefont
  {Georgescu}}, \bibinfo {author} {\bibfnamefont {S.}~\bibnamefont {Ashhab}},\
  and\ \bibinfo {author} {\bibfnamefont {F.}~\bibnamefont {Nori}},\ }\bibfield
  {title} {\bibinfo {title} {Quantum simulation},\ }\href
  {https://doi.org/10.1103/RevModPhys.86.153} {\bibfield  {journal} {\bibinfo
  {journal} {Rev. Mod. Phys.}\ }\textbf {\bibinfo {volume} {86}},\ \bibinfo
  {pages} {153} (\bibinfo {year} {2014})}\BibitemShut {NoStop}%
\bibitem [{\citenamefont {Sch{\"a}fer}\ \emph {et~al.}(2020)\citenamefont
  {Sch{\"a}fer}, \citenamefont {Fukuhara}, \citenamefont {Sugawa},
  \citenamefont {Takasu},\ and\ \citenamefont {Takahashi}}]{Schafer2020}%
  \BibitemOpen
  \bibfield  {author} {\bibinfo {author} {\bibfnamefont {F.}~\bibnamefont
  {Sch{\"a}fer}}, \bibinfo {author} {\bibfnamefont {T.}~\bibnamefont
  {Fukuhara}}, \bibinfo {author} {\bibfnamefont {S.}~\bibnamefont {Sugawa}},
  \bibinfo {author} {\bibfnamefont {Y.}~\bibnamefont {Takasu}},\ and\ \bibinfo
  {author} {\bibfnamefont {Y.}~\bibnamefont {Takahashi}},\ }\bibfield  {title}
  {\bibinfo {title} {Tools for quantum simulation with ultracold atoms in
  optical lattices},\ }\href {https://doi.org/10.1038/s42254-020-0195-3}
  {\bibfield  {journal} {\bibinfo  {journal} {Nature Reviews Physics}\ }\textbf
  {\bibinfo {volume} {2}},\ \bibinfo {pages} {411} (\bibinfo {year}
  {2020})}\BibitemShut {NoStop}%
\bibitem [{\citenamefont {Kaufman}\ \emph {et~al.}(2014)\citenamefont
  {Kaufman}, \citenamefont {Lester}, \citenamefont {Reynolds}, \citenamefont
  {Wall}, \citenamefont {Foss-Feig}, \citenamefont {Hazzard}, \citenamefont
  {Rey},\ and\ \citenamefont {Regal}}]{Kaufman2014}%
  \BibitemOpen
  \bibfield  {author} {\bibinfo {author} {\bibfnamefont {A.~M.}\ \bibnamefont
  {Kaufman}}, \bibinfo {author} {\bibfnamefont {B.~J.}\ \bibnamefont {Lester}},
  \bibinfo {author} {\bibfnamefont {C.~M.}\ \bibnamefont {Reynolds}}, \bibinfo
  {author} {\bibfnamefont {M.~L.}\ \bibnamefont {Wall}}, \bibinfo {author}
  {\bibfnamefont {M.}~\bibnamefont {Foss-Feig}}, \bibinfo {author}
  {\bibfnamefont {K.~R.~A.}\ \bibnamefont {Hazzard}}, \bibinfo {author}
  {\bibfnamefont {A.~M.}\ \bibnamefont {Rey}},\ and\ \bibinfo {author}
  {\bibfnamefont {C.~A.}\ \bibnamefont {Regal}},\ }\bibfield  {title} {\bibinfo
  {title} {Two-particle quantum interference in tunnel-coupled optical
  tweezers},\ }\href {https://doi.org/10.1126/science.1250057} {\bibfield
  {journal} {\bibinfo  {journal} {Science}\ }\textbf {\bibinfo {volume}
  {345}},\ \bibinfo {pages} {306} (\bibinfo {year} {2014})}\BibitemShut
  {NoStop}%
\bibitem [{\citenamefont {Endres}\ \emph {et~al.}(2016)\citenamefont {Endres},
  \citenamefont {Bernien}, \citenamefont {Keesling}, \citenamefont {Levine},
  \citenamefont {Anschuetz}, \citenamefont {Krajenbrink}, \citenamefont
  {Senko}, \citenamefont {Vuletic}, \citenamefont {Greiner},\ and\
  \citenamefont {Lukin}}]{Endres_2016}%
  \BibitemOpen
  \bibfield  {author} {\bibinfo {author} {\bibfnamefont {M.}~\bibnamefont
  {Endres}}, \bibinfo {author} {\bibfnamefont {H.}~\bibnamefont {Bernien}},
  \bibinfo {author} {\bibfnamefont {A.}~\bibnamefont {Keesling}}, \bibinfo
  {author} {\bibfnamefont {H.}~\bibnamefont {Levine}}, \bibinfo {author}
  {\bibfnamefont {E.~R.}\ \bibnamefont {Anschuetz}}, \bibinfo {author}
  {\bibfnamefont {A.}~\bibnamefont {Krajenbrink}}, \bibinfo {author}
  {\bibfnamefont {C.}~\bibnamefont {Senko}}, \bibinfo {author} {\bibfnamefont
  {V.}~\bibnamefont {Vuletic}}, \bibinfo {author} {\bibfnamefont
  {M.}~\bibnamefont {Greiner}},\ and\ \bibinfo {author} {\bibfnamefont {M.~D.}\
  \bibnamefont {Lukin}},\ }\bibfield  {title} {\bibinfo {title} {Atom-by-atom
  assembly of defect-free one-dimensional cold atom arrays},\ }\href
  {https://doi.org/10.1126/science.aah3752} {\bibfield  {journal} {\bibinfo
  {journal} {Science}\ }\textbf {\bibinfo {volume} {354}},\ \bibinfo {pages}
  {1024} (\bibinfo {year} {2016})}\BibitemShut {NoStop}%
\bibitem [{\citenamefont {de~L{\'e}s{\'e}leuc}\ \emph
  {et~al.}(2019)\citenamefont {de~L{\'e}s{\'e}leuc}, \citenamefont {Lienhard},
  \citenamefont {Scholl}, \citenamefont {Barredo}, \citenamefont {Weber},
  \citenamefont {Lang}, \citenamefont {B{\"u}chler}, \citenamefont {Lahaye},\
  and\ \citenamefont {Browaeys}}]{deLeseleuc2019}%
  \BibitemOpen
  \bibfield  {author} {\bibinfo {author} {\bibfnamefont {S.}~\bibnamefont
  {de~L{\'e}s{\'e}leuc}}, \bibinfo {author} {\bibfnamefont {V.}~\bibnamefont
  {Lienhard}}, \bibinfo {author} {\bibfnamefont {P.}~\bibnamefont {Scholl}},
  \bibinfo {author} {\bibfnamefont {D.}~\bibnamefont {Barredo}}, \bibinfo
  {author} {\bibfnamefont {S.}~\bibnamefont {Weber}}, \bibinfo {author}
  {\bibfnamefont {N.}~\bibnamefont {Lang}}, \bibinfo {author} {\bibfnamefont
  {H.~P.}\ \bibnamefont {B{\"u}chler}}, \bibinfo {author} {\bibfnamefont
  {T.}~\bibnamefont {Lahaye}},\ and\ \bibinfo {author} {\bibfnamefont
  {A.}~\bibnamefont {Browaeys}},\ }\bibfield  {title} {\bibinfo {title}
  {Observation of a symmetry-protected topological phase of interacting bosons
  with rydberg atoms},\ }\href {https://doi.org/10.1126/science.aav9105}
  {\bibfield  {journal} {\bibinfo  {journal} {Science}\ }\textbf {\bibinfo
  {volume} {365}},\ \bibinfo {pages} {775} (\bibinfo {year}
  {2019})}\BibitemShut {NoStop}%
\bibitem [{\citenamefont {Barontini}\ and\ \citenamefont
  {Paternostro}(2019)}]{Barontini_2019}%
  \BibitemOpen
  \bibfield  {author} {\bibinfo {author} {\bibfnamefont {G.}~\bibnamefont
  {Barontini}}\ and\ \bibinfo {author} {\bibfnamefont {M.}~\bibnamefont
  {Paternostro}},\ }\bibfield  {title} {\bibinfo {title} {Ultra-cold
  single-atom quantum heat engines},\ }\href
  {https://doi.org/10.1088/1367-2630/ab2684} {\bibfield  {journal} {\bibinfo
  {journal} {New Journal of Physics}\ }\textbf {\bibinfo {volume} {21}},\
  \bibinfo {pages} {063019} (\bibinfo {year} {2019})}\BibitemShut {NoStop}%
\bibitem [{\citenamefont {Kaufman}\ \emph {et~al.}(2012)\citenamefont
  {Kaufman}, \citenamefont {Lester},\ and\ \citenamefont
  {Regal}}]{Kaufman_2012}%
  \BibitemOpen
  \bibfield  {author} {\bibinfo {author} {\bibfnamefont {A.~M.}\ \bibnamefont
  {Kaufman}}, \bibinfo {author} {\bibfnamefont {B.~J.}\ \bibnamefont
  {Lester}},\ and\ \bibinfo {author} {\bibfnamefont {C.~A.}\ \bibnamefont
  {Regal}},\ }\bibfield  {title} {\bibinfo {title} {Cooling a single atom in an
  optical tweezer to its quantum ground state},\ }\href
  {https://doi.org/10.1103/PhysRevX.2.041014} {\bibfield  {journal} {\bibinfo
  {journal} {Phys. Rev. X}\ }\textbf {\bibinfo {volume} {2}},\ \bibinfo {pages}
  {041014} (\bibinfo {year} {2012})}\BibitemShut {NoStop}%
\bibitem [{\citenamefont {Zoller}\ \emph {et~al.}(1995)\citenamefont {Zoller},
  \citenamefont {Maciej}, \citenamefont {Cirac}, \citenamefont {Lewenstein},
  \citenamefont {Cirac},\ and\ \citenamefont {Zoller}}]{Zoller1995}%
  \BibitemOpen
  \bibfield  {author} {\bibinfo {author} {\bibfnamefont {P.}~\bibnamefont
  {Zoller}}, \bibinfo {author} {\bibfnamefont {I.}~\bibnamefont {Maciej}},
  \bibinfo {author} {\bibfnamefont {J.~I.}\ \bibnamefont {Cirac}}, \bibinfo
  {author} {\bibfnamefont {M.}~\bibnamefont {Lewenstein}}, \bibinfo {author}
  {\bibfnamefont {J.~I.}\ \bibnamefont {Cirac}},\ and\ \bibinfo {author}
  {\bibfnamefont {P.}~\bibnamefont {Zoller}},\ }\bibfield  {title} {\bibinfo
  {title} {Master equation for sympathetic cooling of trapped particles},\
  }\href {https://doi.org/10.1103/PhysRevA.51.4617} {\bibfield  {journal}
  {\bibinfo  {journal} {Physical Review A}\ }\textbf {\bibinfo {volume} {51}},\
  \bibinfo {pages} {4617} (\bibinfo {year} {1995})}\BibitemShut {NoStop}%
\bibitem [{\citenamefont {Papenbrock}\ \emph {et~al.}(2002)\citenamefont
  {Papenbrock}, \citenamefont {Salgueiro},\ and\ \citenamefont
  {Weidenm{\"{u}}ller}}]{Papenbrock2002}%
  \BibitemOpen
  \bibfield  {author} {\bibinfo {author} {\bibfnamefont {T.}~\bibnamefont
  {Papenbrock}}, \bibinfo {author} {\bibfnamefont {A.~N.}\ \bibnamefont
  {Salgueiro}},\ and\ \bibinfo {author} {\bibfnamefont {H.~A.}\ \bibnamefont
  {Weidenm{\"{u}}ller}},\ }\bibfield  {title} {\bibinfo {title} {{Rate
  equations for sympathetic cooling of trapped bosons or fermions}},\ }\href
  {https://doi.org/10.1103/PhysRevA.65.043601} {\bibfield  {journal} {\bibinfo
  {journal} {Physical Review A. Atomic, Molecular, and Optical Physics}\
  }\textbf {\bibinfo {volume} {65}},\ \bibinfo {pages} {436011} (\bibinfo
  {year} {2002})}\BibitemShut {NoStop}%
\bibitem [{\citenamefont {Mu\~noz}\ \emph {et~al.}(2020)\citenamefont
  {Mu\~noz}, \citenamefont {Wang}, \citenamefont {Hewitt}, \citenamefont
  {Kowalczyk}, \citenamefont {Sawant},\ and\ \citenamefont
  {Barontini}}]{Munoz_2020_dissipative}%
  \BibitemOpen
  \bibfield  {author} {\bibinfo {author} {\bibfnamefont {J.~M.}\ \bibnamefont
  {Mu\~noz}}, \bibinfo {author} {\bibfnamefont {X.}~\bibnamefont {Wang}},
  \bibinfo {author} {\bibfnamefont {T.}~\bibnamefont {Hewitt}}, \bibinfo
  {author} {\bibfnamefont {A.~U.}\ \bibnamefont {Kowalczyk}}, \bibinfo {author}
  {\bibfnamefont {R.}~\bibnamefont {Sawant}},\ and\ \bibinfo {author}
  {\bibfnamefont {G.}~\bibnamefont {Barontini}},\ }\bibfield  {title} {\bibinfo
  {title} {Dissipative distillation of supercritical quantum gases},\ }\href
  {https://doi.org/10.1103/PhysRevLett.125.020403} {\bibfield  {journal}
  {\bibinfo  {journal} {Phys. Rev. Lett.}\ }\textbf {\bibinfo {volume} {125}},\
  \bibinfo {pages} {020403} (\bibinfo {year} {2020})}\BibitemShut {NoStop}%
\bibitem [{\citenamefont {Thalhammer}\ \emph {et~al.}(2008)\citenamefont
  {Thalhammer}, \citenamefont {Barontini}, \citenamefont {De~Sarlo},
  \citenamefont {Catani}, \citenamefont {Minardi},\ and\ \citenamefont
  {Inguscio}}]{thalhammer}%
  \BibitemOpen
  \bibfield  {author} {\bibinfo {author} {\bibfnamefont {G.}~\bibnamefont
  {Thalhammer}}, \bibinfo {author} {\bibfnamefont {G.}~\bibnamefont
  {Barontini}}, \bibinfo {author} {\bibfnamefont {L.}~\bibnamefont {De~Sarlo}},
  \bibinfo {author} {\bibfnamefont {J.}~\bibnamefont {Catani}}, \bibinfo
  {author} {\bibfnamefont {F.}~\bibnamefont {Minardi}},\ and\ \bibinfo {author}
  {\bibfnamefont {M.}~\bibnamefont {Inguscio}},\ }\bibfield  {title} {\bibinfo
  {title} {Double species bose-einstein condensate with tunable interspecies
  interactions},\ }\href {https://doi.org/10.1103/PhysRevLett.100.210402}
  {\bibfield  {journal} {\bibinfo  {journal} {Phys. Rev. Lett.}\ }\textbf
  {\bibinfo {volume} {100}},\ \bibinfo {pages} {210402} (\bibinfo {year}
  {2008})}\BibitemShut {NoStop}%
\bibitem [{\citenamefont {Fedichev}\ \emph {et~al.}(1996)\citenamefont
  {Fedichev}, \citenamefont {Reynolds},\ and\ \citenamefont
  {Shlyapnikov}}]{Fedichev_1996}%
  \BibitemOpen
  \bibfield  {author} {\bibinfo {author} {\bibfnamefont {P.~O.}\ \bibnamefont
  {Fedichev}}, \bibinfo {author} {\bibfnamefont {M.~W.}\ \bibnamefont
  {Reynolds}},\ and\ \bibinfo {author} {\bibfnamefont {G.~V.}\ \bibnamefont
  {Shlyapnikov}},\ }\bibfield  {title} {\bibinfo {title} {Three-body
  recombination of ultracold atoms to a weakly bound $\mathit{s}$ level},\
  }\href {https://doi.org/10.1103/PhysRevLett.77.2921} {\bibfield  {journal}
  {\bibinfo  {journal} {Phys. Rev. Lett.}\ }\textbf {\bibinfo {volume} {77}},\
  \bibinfo {pages} {2921} (\bibinfo {year} {1996})}\BibitemShut {NoStop}%
\bibitem [{\citenamefont {Caldeira}\ and\ \citenamefont
  {Leggett}(1981)}]{Caldeira_1981}%
  \BibitemOpen
  \bibfield  {author} {\bibinfo {author} {\bibfnamefont {A.~O.}\ \bibnamefont
  {Caldeira}}\ and\ \bibinfo {author} {\bibfnamefont {A.~J.}\ \bibnamefont
  {Leggett}},\ }\bibfield  {title} {\bibinfo {title} {Influence of dissipation
  on quantum tunneling in macroscopic systems},\ }\href
  {https://doi.org/10.1103/PhysRevLett.46.211} {\bibfield  {journal} {\bibinfo
  {journal} {Phys. Rev. Lett.}\ }\textbf {\bibinfo {volume} {46}},\ \bibinfo
  {pages} {211} (\bibinfo {year} {1981})}\BibitemShut {NoStop}%
\bibitem [{\citenamefont {Dann}\ \emph {et~al.}(2019)\citenamefont {Dann},
  \citenamefont {Tobalina},\ and\ \citenamefont
  {Kosloff}}]{Dann_shortcut_2019}%
  \BibitemOpen
  \bibfield  {author} {\bibinfo {author} {\bibfnamefont {R.}~\bibnamefont
  {Dann}}, \bibinfo {author} {\bibfnamefont {A.}~\bibnamefont {Tobalina}},\
  and\ \bibinfo {author} {\bibfnamefont {R.}~\bibnamefont {Kosloff}},\
  }\bibfield  {title} {\bibinfo {title} {Shortcut to equilibration of an open
  quantum system},\ }\href {https://doi.org/10.1103/PhysRevLett.122.250402}
  {\bibfield  {journal} {\bibinfo  {journal} {Phys. Rev. Lett.}\ }\textbf
  {\bibinfo {volume} {122}},\ \bibinfo {pages} {250402} (\bibinfo {year}
  {2019})}\BibitemShut {NoStop}%
\bibitem [{\citenamefont {Pancotti}\ \emph {et~al.}(2020)\citenamefont
  {Pancotti}, \citenamefont {Scandi}, \citenamefont {Mitchison},\ and\
  \citenamefont {Perarnau-Llobet}}]{Pancotti_speedup_2020}%
  \BibitemOpen
  \bibfield  {author} {\bibinfo {author} {\bibfnamefont {N.}~\bibnamefont
  {Pancotti}}, \bibinfo {author} {\bibfnamefont {M.}~\bibnamefont {Scandi}},
  \bibinfo {author} {\bibfnamefont {M.~T.}\ \bibnamefont {Mitchison}},\ and\
  \bibinfo {author} {\bibfnamefont {M.}~\bibnamefont {Perarnau-Llobet}},\
  }\bibfield  {title} {\bibinfo {title} {Speed-ups to isothermality: Enhanced
  quantum thermal machines through control of the system-bath coupling},\
  }\href {https://doi.org/10.1103/PhysRevX.10.031015} {\bibfield  {journal}
  {\bibinfo  {journal} {Phys. Rev. X}\ }\textbf {\bibinfo {volume} {10}},\
  \bibinfo {pages} {031015} (\bibinfo {year} {2020})}\BibitemShut {NoStop}%
\bibitem [{\citenamefont {Scully}\ \emph {et~al.}(2003)\citenamefont {Scully},
  \citenamefont {Zubairy}, \citenamefont {Agarwal},\ and\ \citenamefont
  {Walther}}]{Scully_2009}%
  \BibitemOpen
  \bibfield  {author} {\bibinfo {author} {\bibfnamefont {M.~O.}\ \bibnamefont
  {Scully}}, \bibinfo {author} {\bibfnamefont {M.~S.}\ \bibnamefont {Zubairy}},
  \bibinfo {author} {\bibfnamefont {G.~S.}\ \bibnamefont {Agarwal}},\ and\
  \bibinfo {author} {\bibfnamefont {H.}~\bibnamefont {Walther}},\ }\bibfield
  {title} {\bibinfo {title} {Extracting work from a single heat bath via
  vanishing quantum coherence},\ }\href
  {https://doi.org/10.1126/science.1078955} {\bibfield  {journal} {\bibinfo
  {journal} {Science}\ }\textbf {\bibinfo {volume} {299}},\ \bibinfo {pages}
  {862} (\bibinfo {year} {2003})}\BibitemShut {NoStop}%
\bibitem [{\citenamefont {Dillenschneider}\ and\ \citenamefont
  {Lutz}(2009)}]{Dillenschneider_2009}%
  \BibitemOpen
  \bibfield  {author} {\bibinfo {author} {\bibfnamefont {R.}~\bibnamefont
  {Dillenschneider}}\ and\ \bibinfo {author} {\bibfnamefont {E.}~\bibnamefont
  {Lutz}},\ }\bibfield  {title} {\bibinfo {title} {Energetics of quantum
  correlations},\ }\href {https://doi.org/10.1209/0295-5075/88/50003}
  {\bibfield  {journal} {\bibinfo  {journal} {{EPL} (Europhysics Letters)}\
  }\textbf {\bibinfo {volume} {88}},\ \bibinfo {pages} {50003} (\bibinfo {year}
  {2009})}\BibitemShut {NoStop}%
\bibitem [{\citenamefont {Ro\ss{}nagel}\ \emph {et~al.}(2014)\citenamefont
  {Ro\ss{}nagel}, \citenamefont {Abah}, \citenamefont {Schmidt-Kaler},
  \citenamefont {Singer},\ and\ \citenamefont {Lutz}}]{Rossnagel_2014}%
  \BibitemOpen
  \bibfield  {author} {\bibinfo {author} {\bibfnamefont {J.}~\bibnamefont
  {Ro\ss{}nagel}}, \bibinfo {author} {\bibfnamefont {O.}~\bibnamefont {Abah}},
  \bibinfo {author} {\bibfnamefont {F.}~\bibnamefont {Schmidt-Kaler}}, \bibinfo
  {author} {\bibfnamefont {K.}~\bibnamefont {Singer}},\ and\ \bibinfo {author}
  {\bibfnamefont {E.}~\bibnamefont {Lutz}},\ }\bibfield  {title} {\bibinfo
  {title} {Nanoscale heat engine beyond the carnot limit},\ }\href
  {https://doi.org/10.1103/PhysRevLett.112.030602} {\bibfield  {journal}
  {\bibinfo  {journal} {Phys. Rev. Lett.}\ }\textbf {\bibinfo {volume} {112}},\
  \bibinfo {pages} {030602} (\bibinfo {year} {2014})}\BibitemShut {NoStop}%
\end{thebibliography}%
	
\end{document}